\documentclass[aps,prl,twocolumn,floatfix]{revtex4-2}
\usepackage{enumitem}
\usepackage{hyperref}
\usepackage{graphicx}
\usepackage{xcolor}
\usepackage{soul}
\usepackage{svg}
\usepackage{booktabs}

\begin{document}

\title{Lunar Silicon Cavity}
\author{Jun Ye}
 \email{ye@jila.colorado.edu}
\author{Zoey Z. Hu}
\author{Ben Lewis}
\affiliation{
    JILA, National Institute of Standards and Technology and University of Colorado, Boulder, Colorado 80309-0440, USA
}%

\author{Wei Zhang}
\affiliation{%
Jet Propulsion Laboratory, NASA and Caltech, USA
}%

\author{Fritz Riehle, Uwe Sterr}
\affiliation{%
Physikalisch-Technische Bundesanstalt, Bundesallee 100, 38116 Braunschweig, Germany
}%

\author{Yiqi Ni}
\email{yiqi@lunetronic.com}

\author{Julian Struck}
\affiliation{%
Lunetronic Inc. San Francisco, California 94109, USA
}%

\date{\today}

\begin{abstract}
The Moon's permanently shadowed regions (PSRs) are among the coldest places in the Solar System and are expected to become key landing sites for upcoming international space agency missions. Their proximity to peaks of perpetual solar power and potential resource richness makes them prime candidates for lunar exploration and future Moon bases. Here we propose to deploy a passive, ultrastable optical resonator in these regions that will enable laser systems with unprecedented phase-coherence. The unique physical environment of lunar PSRs greatly benefits the construction of a cryogenic monolithic silicon cavity that exhibits low $10^{-18}$ thermal noise-limited stability and coherence time exceeding 1 minute, more than a decade better than the current best terrestrial system. Such a stable laser will form an enabling infrastructure for quantum technology in space to serve many applications, including establishing a lunar time standard, building long-baseline optical interferometry, distribution of stable optical signals across networks of satellites, testing general relativity and gravitational physics, and forming the backbone for space-based quantum networks.
\end{abstract} 
\maketitle

Aspirations for developing and deploying space-borne quantum technologies can in large part be facilitated by having access to ultrastable lasers that are key to driving and interconnecting optically active quantum systems. On the ground, stable optical local oscillators play many versatile roles in, for example, state-of-the-art optical atomic clocks~\cite{matei2017, Oelker2019,bothwell2025transportableYb, marshall2025high, hobson2020strontium, tofful2024yb, koller2017transportable}, precision test of fundamental physics~\cite{Braxmaier2002CryogenicOpticalResonator, Sanjuan2019LongTermStableOpticalCavity, Bothwell2022,Pikovski2012,Mehlstaubler2018}, long-baseline optical interferometry including gravitational wave detection~\cite{Barontini2025DetectingMilliHzGWs}, and optical networks for both classical and quantum information~\cite{nardelli2025phasestable, riehle2017optical}. Cryogenic silicon cavities enable the best performing frequency-stable lasers, with fractional frequency stability of low $10^{-17}$ and an ultralow linear drift rate of mid $10^{-20}$/s~\cite{Lee2026, kessler2012cryogenic}. Permanently shadowed regions (PSRs) on the Moon offer low-temperature and ultra-high vacuum conditions in combination with exceptionally low vibrational noise. This environment is ideal for supporting an ultrastable optical cavity with performance surpassing the best terrestrial systems. Having access to a master optical oscillator frequency stabilized to the lunar cavity forms the basic infrastructure for a range of space-borne experiments, with a cascade of satellites housing either secondary lasers or atomic quantum systems networked together via phase stable optical links. 

Lunar PSRs are the chosen landing sites for the NASA-led Artemis mission and other international missions due to their likely resource richness that includes water ice, carbon dioxide and helium-3~\cite{Paige2010DivinerRegion}, as well as continuous solar power at nearby peaks of eternal light~\cite{Bussey1999IlluminationPole,Bickel2021PeeringLearning,BenBussey2003PermanentPoles}. However, landing and navigating near PSRs face significant technical challenges. The low Sun elevation angles and extended shadows limit optical and terrain-relative navigation capability, making precision positioning, navigation, and timing (PNT) essential for safely landing payloads and crewed missions. 

We propose to construct a cryogenic silicon cavity located in a PSR to take advantage of the uniquely beneficial physical conditions for such an optical reference cavity. PSRs’ ambient cryogenic temperature and easy access to radiative cooling from deep space permit a simple and passive cooling strategy to reach a zero of the silicon cavity's thermal expansion coefficient at 17~K. The extreme thermal stability from isolation of solar radiation will enable exceptional long term frequency stability of the optical oscillator. The lunar seismic noise is orders of magnitude lower than in a terrestrial laboratory environment, facilitating robust performance for an extended cavity length to scale down the fundamental Brownian thermal noise contribution. The high vacuum environment of PSRs also eases the construction requirement for a cavity chamber. Overall, once the silicon cavity material is transported to the Moon, the final system engineering will be straightforward to implement and yet with far greater performance prospects.

A highly phase-coherent lunar master laser can serve many important tasks for emerging scientific and technological explorations via space-borne experiments~\cite{Kolkowitz2016,Turyshev2025,Zhu2025TransportableApplications,Kopeikin2024LunarRelativity,Lezius2016Space-borneMetrology}.
In the absence of atmospheric perturbation, it will be much easier to establish a lunar-space optical frequency/phase transfer link than starting with a stable oscillator on Earth. It is expected that while Earth-satellite optical links may demonstrate $10^{-18}$ performance, they will take $>1000$~s to average to that level~\cite{Caldwell_OpticalTimeTransfer,Shen_OpticalTimeTransfer} which is not fast enough to transfer laser coherence. However, the lunar laser's phase stability can be transferred with high fidelity to secondary systems on board of various satellite clusters/constellations, enabling the construction of long baseline space-borne optical interferometers~\cite{Berceau2016,Derevianko2022}. Each satellite with a laser onboard can have its optical field phase-locked to the lunar master laser, greatly simplifying on-board frequency control systems for networks of satellites used for classical or quantum communications, as well as for navigation and flight formation needs. State-of-the-art atomic clocks rely on high-performance and robust local oscillators. The stability of the clock is often determined by the phase coherence of the driving laser. Housing a stable optical reference cavity onboard a satellite to accompany its atomic payload remains a technically sophisticated task. With phase-stable optical links established between satellites and the lunar laser base, any onboard atomic systems can be quickly turned into optical atomic clocks, and benefit from the unmatched phase coherence of the lunar laser. 

Another natural outcome following the lunar stable laser is the foundation for an optical atomic clock for lunar time standard~\cite{Bourgoin2025LunarTimescale}. The lunar silicon cavity's exceptional long-term frequency stability is, by itself, sufficient for a wide range of applications. With performance below $10^{-15}$ at a 1 day timescale, the silicon cavity can easily provide a reference for any PNT requirements, and form the backbone of Lunar Coordinated Time (LTC)~\cite{Bourgoin2025LunarTimescale, Ashby2024AMoon, FiengaRambauxSosnica2024Moonlight}. However, to gain even greater long-term stability, an atomic standard could be added. Only very low-frequency steering would be required, allowing the atomic standard to be located anywhere within the network connected to the cavity. 

This proposal thus addresses a critical need for establishing a standalone LTC serving as the cornerstone of the future PNT infrastructure of the Moon. The success of this mission will mark a historic milestone, demonstrating humanity’s capability to build fundamental infrastructure for quantum technologies on another celestial body, alongside establishing a permanent presence on the Moon. This achievement can have profound implications for future Mars and deep-space missions, where terrestrial timing infrastructure is nearly impossible to access, making a standalone local time reference the only viable solution~\cite{Burt2021DemonstrationSpace}.

\section*{PSR environment characteristics}

The Moon’s spin axis is nearly perpendicular to its orbital plane around the Sun, causing craters near lunar poles to remain in permanent shadow. These regions, known as PSRs, have eluded sunlight for billions of years. Combined with the lack of internal heating from the Moon’s inactive core and low residual heat, PSRs are among the coldest known locations in the entire Solar System. Over time, large amounts of volatile compounds, such as water ice, helium, and carbon dioxide, have been trapped within PSRs. Mining and utilizing these resources is critical for ongoing deep space exploration, enabling the Moon to serve as a sustainable commercial human base, a refueling outpost, and a testbed for technologies before deploying them to other planets such as Mars.

However, landing payloads near PSRs presents major challenges. In the absence of a terrestrial-like global navigation satellite system, vision-based landing remains one of the few reliable navigation methods. Yet, because the solar incidence angle near PSRs is nearly parallel to the surface, objects cast extremely long shadows, creating significant obstacles for precision lunar landing. A local timescale will solve this problem.

We summarize that the PSRs environment (Table I) is ideal for deploying a cryogenic silicon cavity-based timescale, close to future permanent human bases, utilizing the naturally maintenance-free conditions of PSRs. 

\begin{figure}[t]
\centering
\includegraphics[width=8.7cm]{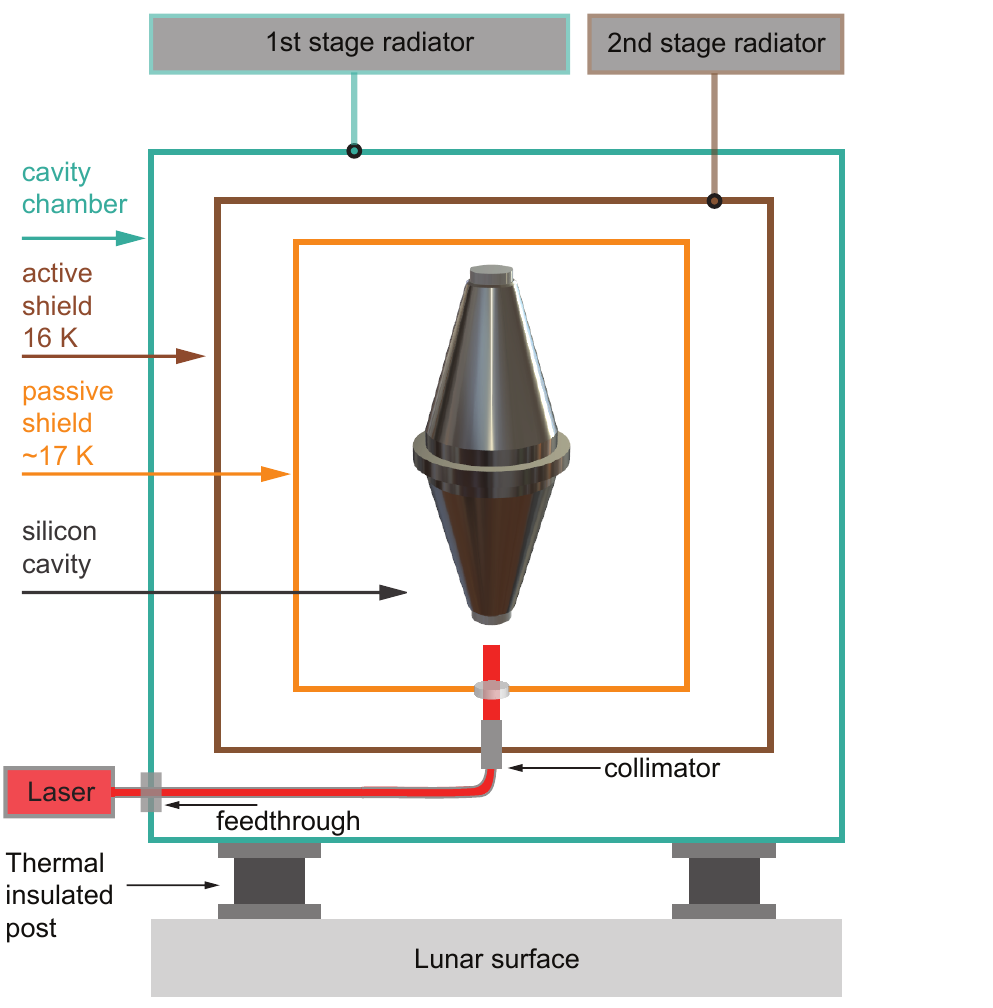}
\caption{Conceptual design of a cryogenic silicon cavity based on radiative cooling. The whole system supported by the thermal insulated posts is constructed on the lunar surface. Two radiators, both facing the deep space, provide cooling for an outer radiation shield and an inner actively controlled thermal shield for a 17~K operating temperature where the silicon thermal expansion coefficient is zero. A continuous-wave laser is sent into the cavity chamber via a fiber-optic feedthrough or free-space motorized mirrors and coupled to the silicon cavity with a collimator.}
\label{fig:1}
\end{figure}

\subsection*{Temperature}

\nocite{Johnson1972VacuumSurface,Lorenz2018EmpiricalSurfaces}
The thermal environment of the south polar PSRs has been monitored for over a decade through remote sensing by NASA’s Lunar Reconnaissance Orbiter, ranging from about 20~K in winter to 60~K in summer~\cite{Williams2019SeasonalMoon}. Seasonal variations are primarily driven by the Moon’s 1.5$^\circ$ axial tilt relative to the Earth-Sun orbital plane. Owing to the slow motion of this tilt and the extreme isolation of PSRs, temperature changes are both gradual and highly predictable, approximately 50 mK per day.

\subsection*{Radiative cooling}

For cooling and thermal stabilization of the silicon cavity, we will engineer access to deep space radiative cooling. This passive cooling method is free of vibration and cryogen and has been widely used in space missions, such as James Webb Space Telescope ~\cite{JWST2022,Menzel_2023} and far-side seismic suite~\cite{FFS_2024}. Figure~\ref{fig:1} illustrates the conceptual design of the cavity system on the lunar surface. Similar to systems currently operated in the lab, three layers of thermal shields are provided for the silicon cavity. The cavity chamber connected to the first radiator is cooled down to 30\textendash40~K. The active shield attached to the second radiator is approximately 16~K, controlled by an active temperature servo. To accommodate the PSR's temperature cycle of 20\textendash60~K from winter to summer, a thermal switch~\cite{FFS_2024} will be used to adjust the first radiator's cooling power, to keep the heating power needed for the active shield below 0.25~W. The passive shield provides a long time constant for thermal equilibrium of the silicon cavity, which is stabilized at the zero-crossing of the coefficient of thermal expansion at approximately 17~K. This fully passive cooling strategy relies on the natural cryogenic environment of the Moon and deep space (2.7~K).

\subsection*{Vacuum}

The Moon’s weak gravity prevents it from retaining a substantial atmosphere, resulting in a naturally maintained ultra-high vacuum environment. Direct measurements from the Apollo missions recorded surface neutral particle pressure of about $10^{-6}$~Pa during the lunar day and $10^{-10}$~Pa at night~\cite{Johnson1972VacuumSurface}. Since the primary source of residual gas on the Moon is solar wind interaction, the vacuum within PSRs is expected to be even lower. This inference is supported by the persistence of volatile compounds, such as water ice, that have remained stable in PSRs cold traps for billions of years.

\subsection*{Ground Vibration}

The absence of a lunar atmosphere eliminates above-ground acoustic noise, and the Moon exhibits an exceptionally low seismic background compared to Earth due to its minimal tectonic activity. Multiple Apollo missions directly measured the lunar seismic background~\cite{NASA_Apollo11Seismic,Latham1970,Nakamura1982,Nunn2021,nunn2020lunar,Watkins1972ApolloExperiment}. In the absence of moonquakes, the ambient ground vibration was found to be below the seismometer noise floor, approximately 0.3~nm at 1~Hz, significantly quieter than any known site on Earth~\cite{Latham1970}. Moon ground vibration can be caused by four major events, deep moonquake, shallow moonquake, thermal moonquake, and meteor impacts~\cite{Toksoz1974,nunn2020lunar2}. Thermal moonquakes are the most common, caused by thermal expansion and contraction between lunar day and night, where surface temperatures can fluctuate by up to 300~K. The average amplitude of thermal moonquakes is about 0.6~nm, with peak values around 6~nm, substantially quieter than a typical Earth-based scientific laboratory ~\cite{Duennebier1974ThermalMoonquakes}. Because the thermal environment within PSRs is far more stable than at the Apollo landing sites, significantly fewer thermal moonquakes are expected inside PSRs. Deep and shallow moonquakes may be induced by tidal stresses from Earth, while meteoroid impacts can occasionally generate localized seismic disturbances~\cite{Majstorovic2025ModelingForcing,Latham1970MoonquakesMeteoroidsLunarInterior}. In general, the Moon’s seismic noise background is about four orders of magnitude lower than even the quietest continental sites on Earth~\cite{Lorenz2018EmpiricalSurfaces}, making it an ideal environment for silicon-cavity-based optical clocks.

The combined conditions of the stable cryogenic thermal environment of the PSR, ultra-high vacuum, and intrinsically low ground vibration create an ideal natural infrastructure to deploy silicon-cavity-based optical clocks as an independent reference for lunar time, as well as other quantum instruments in the future. We propose a low-size, weight, and power (SWaP) architecture that requires minimal to zero active thermal management to realize an ultra-stable timing node for synchronizing lunar assets.

\begin{table}[h]
\begin{center}
\caption{PSR Environment. The temperature range corresponds to remote measurements of a Haworth Crater site. The vacuum pressure is projected based on the Apollo mission data taken in lunar equatorial regions~\cite{Johnson1972VacuumSurface}. The vacuum pressure is expected to be significantly lower in PSRs. The PSR ground vibration is evaluated at Fourier frequency of 1~Hz, with the maximum vibration level associated with artificial impacts of Saturn IVB rocket stages during Apollo~\cite{Lorenz2018EmpiricalSurfaces}.\label{tab:1}} 
\begin{tabular}{@{}c@{\hskip2em}ccc@{}} 
 \toprule
Property & Average & Min & Max \\ \midrule
Temperature (K) & $<0.05$/day drift & $<20$ & 60 \\ 
Vacuum (Pa) & $<10^{-10}$ & & \\ 
Vibration (m/s$^{2}$)& $<1.2\times10^{-8}$ & $<1.2\times10^{-8}$ &$3\times10^{-6}$ \\
 \bottomrule
\end{tabular}
\end{center}
\end{table}

\section*{Silicon cavity properties}

We now connect the favorable environmental conditions of PSRs to the physical properties of silicon cavities to highlight the potential for improving their frequency stability beyond what is available in the ground laboratory today. The low vibration background will reduce high frequency noise associated with the optical cavity. Thermal stability and abundance of the cooling power provide long term stability at zero crossing of the silicon crystal's thermal expansion coefficient. Considering the high vacuum surroundings at PSRs, only a modest shielding chamber is needed to protect the silicon cavity and ensure that the background pressure fluctuation within the optical cavity is below $10^{-10}$~Pa. 

We illustrate the tradeoff between performance and system size by comparing two designs. A 21~cm cavity follows the current design of silicon cavities in the lab. A longer cavity of 50~cm length will be more sensitive to the vibration noise, but the PSR's low vibration environment significantly reduces this concern. The longer 50~cm cavity reduces the fundamental Brownian thermal noise of the mirror coatings and enlarges the optical mode area. For ultimate performance, we can also use mirror coatings based on stacks of crystalline GaAs/AlGaAs for lower thermal noise than conventional dielectric mirrors. Even larger cavities are certainly possible and would further improve performance, but the considered cavities already surpass current records by up to 30$\times$.

Table~\ref{tab:2} shows the thermal noise floor of these cavity designs, which are all below $10^{-17}$, a significant improvement over the state-of-the-art performance of $\sim3\times10^{-17}$ for terrestrial-based cryogenic silicon cavities. We assume the use of one plano mirror and one with a 10~m radius of curvature. Of course, the overall performance of the lunar cavity will also include contributions from other noise sources, including optical beam propagation stability, intracavity gas/vacuum noise, residual amplitude noise in frequency modulation, and so on. However, a stable environment offered by PSRs will correspondingly reduce the magnitude of all these noise contributions.

\begin{table}[h]
\begin{center}
\caption{Thermal noise levels for different cavity designs. Longer cavities and crystalline coatings reduce the noise level.
\label{tab:2}} 
\begin{tabular}{@{}l@{\hskip 1em}lr@{}} \toprule
\multicolumn{2}{c}{Cavity Design} & \\ \cmidrule(r){1-2}
Length & Coating &\hskip 1.5em Thermal noise\\ \midrule
21 cm& Conventional &$9\times10^{-18}$ \\
21 cm& Crystalline &$2\times10^{-18}$ \\
50 cm& Conventional &$3\times10^{-18}$ \\
50 cm& Crystalline &$8\times10^{-19}$ \\ \bottomrule
\end{tabular}
\end{center}
\end{table}

\subsection*{Thermal Management}
Silicon is uniquely suited as an optical cavity material because it has a very stable crystal structure, with a high mechanical quality factor, and most importantly has a zero-crossing of thermal expansion at 17~K. Operating at this temperature allows long-term stability due to insensitivity of the cavity to temperature changes.
To maintain the silicon cavity at a 17~K operating temperature, we estimate the thermal load based on the following conditions. The surface of the cavity chamber, active shield, and passive shield are gold-coated to ensure low emissivity (5$\%$) between different layers. Each layer may be mounted using high strength and low thermal conductivity cables, e.g. Kevlar fiber~\cite{Bugby2003}. The whole system is supported by thermally insulated posts. The coupling to external blackbody radiation is minimized by using a fiber-optic feedthrough. The use of the two separate radiators, as outlined in Fig.~\ref{fig:1}, makes it practical to support a silicon cavity with relatively long length (50~cm and biconical as shown in Fig.~\ref{fig:1}). The size of the first radiator is 10\textendash20~m$^2$, and the second radiator is 1\textendash10~m$^2$. Note that this radiator design includes a conservative margin of 50$\%$ higher heating load than a realistic estimate.

\subsection*{Seismic Noise}

\begin{figure}[t]
    \centering
    \includegraphics[width=8.7cm]{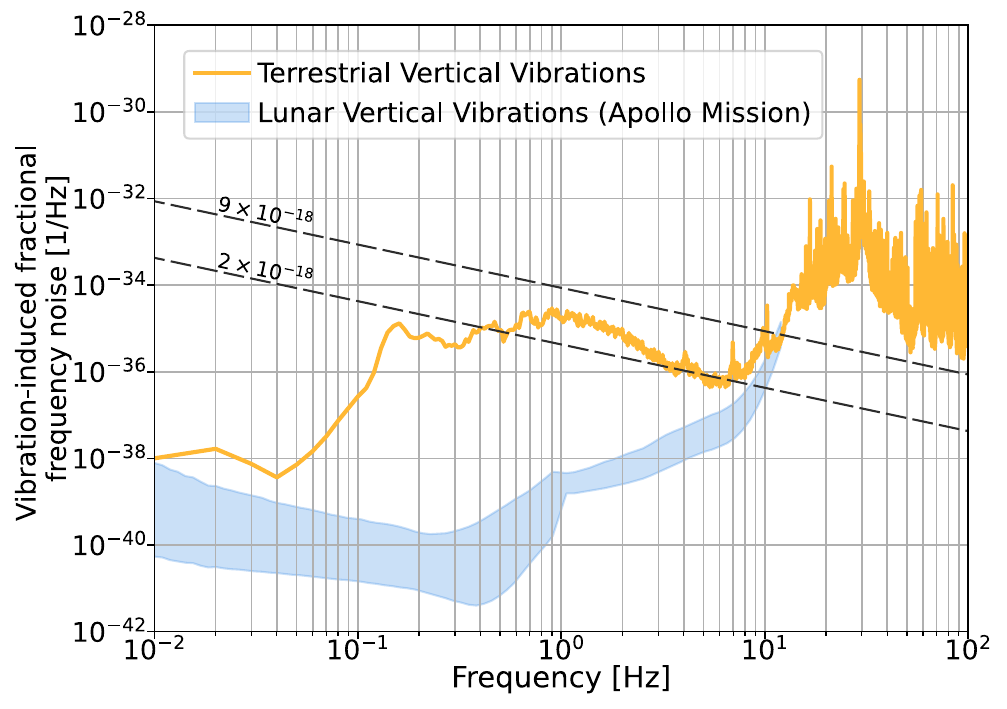}
    \caption{Comparison of terrestrial and lunar vibration-induced fractional frequency noise. The blue shaded region denotes lunar seismic vibration noise in the vertical direction) from the Apollo missions, scaled by the measured vertical acceleration sensitivity of the 21~cm cavity ($1.5\times 10^{-11}/g$), and spanning the 10th to 90th percentile range. These data were constructed and analyzed by~\cite{Nunn2021} from multiple Apollo stations operated in both flat-response and peak-mode seismometer configurations. The orange trace shows an example of typical terrestrial laboratory vibration conditions with considerable vibration control effort. The dashed lines indicate the thermal noise levels for the 21~cm cavity with conventional and crystalline coatings respectively.}
    \label{fig:2}
\end{figure}

The most recent 21~cm silicon cavity has measured acceleration sensitivities of $1\times 10^{-10}/g$ and $3\times 10^{-11}/g$ for two orthogonal horizontal directions, and $1.5\times 10^{-11}/g$ for the vertical direction~\cite{Lewis2026}. Here, $g = 9.8~$m/s$^2$ is the gravitational acceleration on the surface of the Earth. We only have data for the Moon's vibrational noise in the vertical direction, and we expect the horizontal vibrational noise to be much lower than the vertical noise. We can compare the expected vibration-induced noise for a lunar cavity and a terrestrial cavity that are mounted on an actively vibration-isolated table, see Fig.~\ref{fig:2}. The Moon's passive noise, with no isolation, is less than that of an active isolation system on Earth. If horizontal noise is comparable to vertical noise, its effect would be larger due to the higher horizontal sensitivity; nevertheless, it would still be below the $1\times10^{-18}$ level for frequencies below 1~Hz. The 50~cm cavity would have a higher vibration sensitivity, potentially by a factor of 2.5, placing the vibration noise-induced fractional frequency fluctuation around $1\times10^{-18}$ at 1~Hz. At frequencies below 0.1~Hz the vibration contribution would be even less, and even a 1~m or larger cavity could remain thermal-noise limited. Because the cavity is passively cooled, the cooling system does not induce vibrations, as would happen on Earth. Overall, we expect the vibration-induced noise to be below the fundamental thermal noise of the cavity.

\subsection*{Vacuum Pressure}
Residual gas within the spacer hole modifies the effective optical path length through pressure- and temperature-dependent refractive index variations. In the terrestrial lab, refractive index fluctuations in air have been well characterized with the Edlén equation~\cite{Edlen1966RefractiveIndexAir}, which allows the estimation of the vacuum stability required to reach the expected thermal noise limited performance. At 17~K, the fractional frequency varies with pressure $\sim $4$\times10^{-8}$/Pa, assuming that the refractivity per unit density of the residual gas is similar to that of air. For the recent 21~cm terrestrial cavity operated at 17~K, a steady-state pressure of $10^{-7}$~Pa is reached with active pumping, and its fluctuation is estimated to be $\leq$1$\times10^{-10}$~Pa for a stability limit of 5$\times10^{-18}$.

\begin{figure*}[t]
\centering
\includegraphics[width=17.8cm]{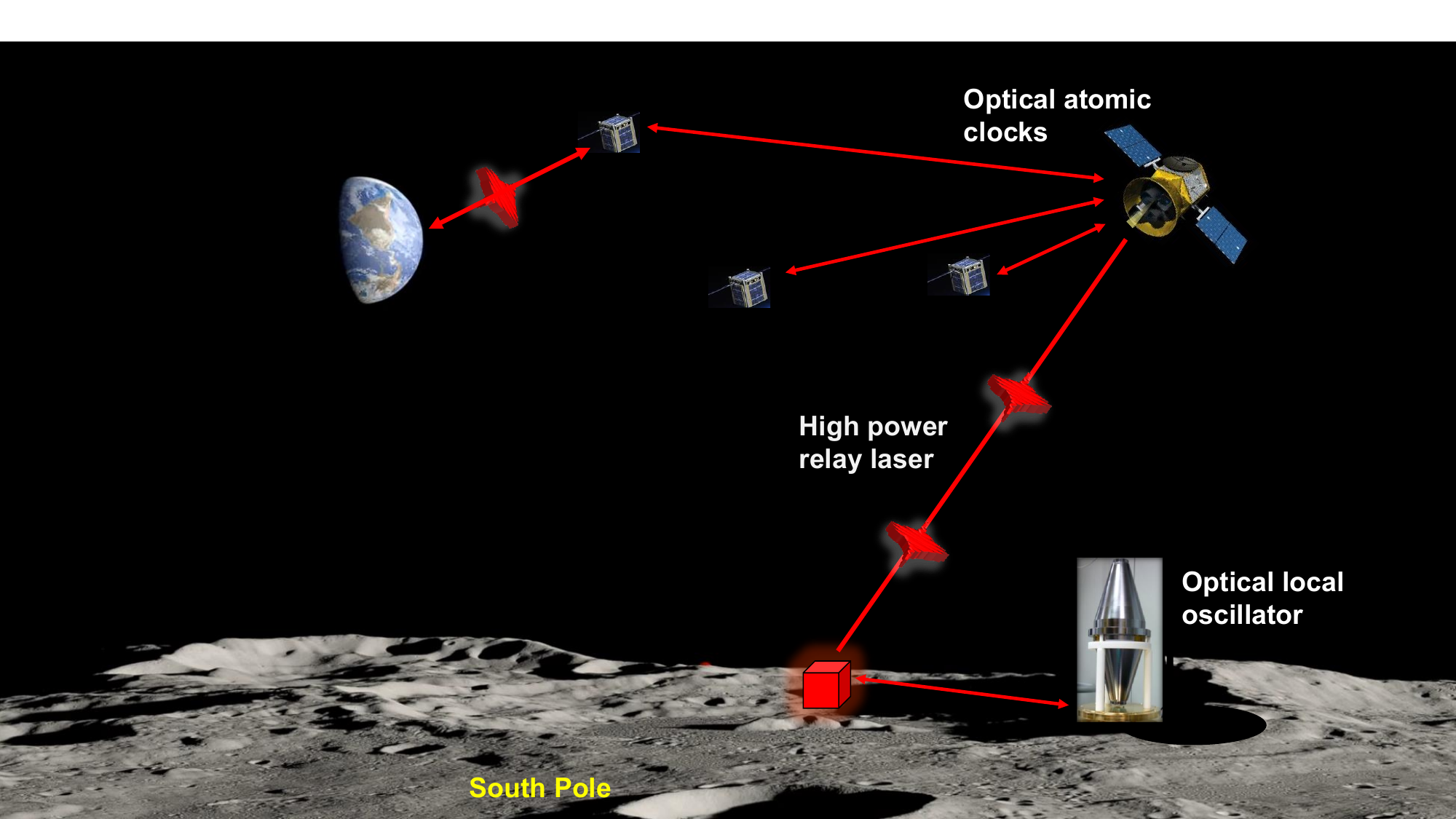}
\caption{Application of a lunar PSR silicon cavity as the basic infrastructure for lunar time scale, Earth-Moon optical communication, satellite-based space interferometry and imaging, and networking to Earth-bound time scale. Lunar PSR background image produced by NASA's visualization studio~\cite{NASA_visualization}.}
\label{fig:3}
\end{figure*}

In contrast, lunar PSRs have naturally extensive cold surfaces that passively suppress the overall vacuum levels and minimal diurnal variation~\cite{Paige2010DivinerRegion,Siegler2015EvolutionStability,Williams2019SeasonalMoon2}. As mentioned previously, direct measurements from the Apollo missions reported a surface neutral particle pressure of 1$\times10^{-10}$~Pa level at night~\cite{Johnson1972VacuumSurface}. These data, not collected within PSR regions, provide a conservative upper bound for expected local pressure level and temporal pressure activities: some of the gases probably condense/adhere to the significantly colder surfaces in the PSRs and their thermal stability guarantees pressure stability since there is no gas desorption cycle related to illumination. Therefore, the local pressure and temporal fluctuations in the PSRs are expected to remain well below the level required thermal noise limited stability performance without the need for an active pump. There are obvious risk factors that need to be mitigated, including the shielding of lunar dust, ionizing radiation, and micrometeorites. 



\section*{Applications of the lunar silicon cavity/Outlook}
A wide range of applications, from fundamental physics to practical communication and navigation needs, will be enabled by a lunar silicon-cavity-based laser system. As shown in Fig.~\ref{fig:3}, with a simple heterodyne beat and optical phase locking setup, the silicon cavity's optical stability can be transferred to a high power laser located on the ridge of PSRs with permanent solar energy, and subsequently connected to an array of satellites that house atomic standards, optical interconnects, interferometric components, and long-baseline capabilities to reach back to Earth.

This stable laser will allow for establishment of a lunar optical atomic clock. The ultrastable silicon cavity already provides sufficiently high stability, with performance that could surpass a maser~\cite{Milner2019OpticalTimescale}. The free-running cavity frequency can provide a stand-alone frequency/phase reference to many demanding applications. If steered by an atomic standard on a regular basis with relatively low duty cycles, the system will provide an optical time scale~\cite{Milner2019OpticalTimescale,Yao2019OpticalClockTimeScale,Hachisu2018MonthsLongTimeScale,Formichella2024YearLongOpticalTimeScale,Zhang2023SrOpticalTimeScaleNTSC}. The atomic standard can be located on the Moon, or on a satellite, with an optical link between the Moon and the satellite, to provide maximum flexibility. Together with the optical frequency comb technology, multiple atomic standards may take advantage of this lunar infrastructure. Such a network of frequency steered and interconnected satellites will greatly aid communication and navigation needs between Earth and the Moon~\cite{Ely2025BenefitSpaceClocksDSN,Berceau2016}. Since a precise time reference underpins the PNT, this network will strengthen the resilience of the lunar mission by enabling networked communications, the collection of correlated scientific data and safe landing, even during long holdover periods without terrestrial signals~\cite{lunanet2020}. 

Ultrastable space-based optical oscillators will also become widely available. Each satellite no longer needs to be equipped with its own cavity-based frequency stabilization device. The lunar silicon cavity can serve as the common reference for connecting satellites, each housing a receiver and potentially a phase locked loop to track the incoming optical phase. A two-way optical frequency/phase transfer can be established between pairs of satellites, as well as between the Moon and satellites. 
This infrastructure will thus play a valuable role in establishing a space-based long baseline optical interferometer for deep space imaging, astronomical observation, and gravitational wave detection. With the Moon's orbit determined at the few-millimeter level~\cite{Battat_2009}, this stable gravitational system also enables unique tests of general relativity. The exquisite silicon cavity at this unique location can enable investigations of fundamental interaction of the cavity itself, such as for mHz gravitational wave detection \cite{Barontini2025DetectingMilliHzGWs} or observation of the effects of dark matter on solids \cite{Wcislo_Dark_Matter_2016}.
The distance between Earth and the Moon is only one light-second. The silicon cavity's optical coherence time far exceeds this time scale, and the ultrastable optical frequency reference, as well as the atomic standard-steered lunar optical time scale, can reach geocentric satellites. Atmospheric perturbations may limit the short-term performance of links between satellites and Earth, but this network would allow long-term clock comparisons among national metrology labs across continents.

We thank J. L. Hall, D. Lee, I. M. McKinley, and J. K. Thompson for technical discussions and useful feedback on the manuscript.
The work is supported by National Institute of Standards and Technology, NSF, Vannevar Bush Faculty Fellowship, the Moore/Simons Foundation, and Jet Propulsion Laboratory’s Strategic University Research Partnerships program under a contract with NASA (80NM0018D0004). F.R. and U.S. acknowledge support from the Deutsche Forschungsgemeinschaft (DFG) under Germany’s Excellence Strategy EXC-2123 QuantumFrontiers (Project No. 390837967)
Y. Ni and J. Struck are with Lunetronic, which is designing future quantum infrastructure inside lunar permanently shadowed regions (PSRs). The other authors declare no competing interests.

\bibliography{Ye_lunar_cavity_arXiv}

@article{matei2017,
  author  = {Matei, D.G. and others},
  title   = {1.5 $\mu$m lasers with sub-10 {mHz} linewidth},
  journal = {Phys. Rev. Lett.},
  volume  = {118},
  pages   = {263202},
  year    = {2017},
}

@article{kessler2012cryogenic,
    author={Kessler, T. and others},
    title={A sub-40-{mHz}-linewidth laser based on a silicon single-crystal optical cavity},
    journal={Nature Photonics},
    year={2012},
    month={Oct},
    day={01},
    volume={6},
    number={10},
    pages={687-692},
    issn={1749-4893},
    doi={10.1038/nphoton.2012.217},
    url={https://doi.org/10.1038/nphoton.2012.217}
}

@article{hobson2020strontium,
  title        = {A strontium optical lattice clock with $1\times10^{-17}$ uncertainty and measurement of its absolute frequency},
  author       = {Hobson, Richard and others},
  journal      = {Metrologia},
  volume       = {57},
  number       = {6},
  pages        = {065026},
  year         = {2020},
  doi          = {10.1088/1681-7575/abb530},
  url          = {https://doi.org/10.1088/1681-7575/abb530},
}

@article{tofful2024yb,
  title        = {$^{171}${Yb}$^+$ optical clock with $2.2\times10^{-18}$ systematic uncertainty and absolute frequency measurements},
  author       = {Tofful, Alexandra and others},
  journal      = {Metrologia},
  volume       = {61},
  number       = {4},
  pages        = {045001},
  year         = {2024},
  doi          = {10.1088/1681-7575/ad53cd},
  url          = {https://iopscience.iop.org/article/10.1088/1681-7575/ad53cd},
}

@article{marshall2025high,
  title        = {High-Stability single-ion clock with $5.5\times10^{-19}$ systematic uncertainty},
  author       = {Marshall, Mason C. and others},
  journal      = {Phys. Rev. Lett.},
  volume       = {135},
  number       = {3},
  pages        = {033201},
  year         = {2025},
  doi          = {10.1103/PhysRevLett.135.033201}
}

@article{bothwell2025transportableYb,
  title        = {Deployment of a transportable {Yb} optical lattice clock},
  author       = {Bothwell, Tobias and others},
  journal      = {Opt. Lett.},
  volume       = {50},
  number       = {2},
  pages        = {646--649},
  year         = {2025},
  doi          = {10.1364/OL.543310},
  url          = {https://doi.org/10.1364/OL.543310},
}

@article{koller2017transportable,
  title        = {Transportable optical lattice clock with $7\times10^{-17}$ uncertainty},
  author       = {Koller, S.~B. and others},
  journal      = {Phys. Rev. Lett.},
  volume       = {118},
  number       = {7},
  pages        = {073601},
  year         = {2017},
  doi          = {10.1103/PhysRevLett.118.073601},
  url          = {https://doi.org/10.1103/PhysRevLett.118.073601},
}

@MISC{nardelli2025phasestable,
  author       = {Nicholas V. Nardelli and others},
  title        = {Phase-stable optical fiber links for quantum network protocols},
  howpublished = {arXiv [Preprint]},
  year         = {2025},
  note         = {\url{https://doi.org/10.48550/arXiv.2510.16230} (Accessed 22 April 2026)}
}

@misc{FiengaRambauxSosnica2024Moonlight,
  title         = {Lunar references systems, frames and time-scales in the context of the {ESA} programme moonlight},
  author        = {Fienga, Agn{\`e}s and others},
  howpublished = {arXiv [Preprint]},
  year         = {2024},
  note         = {\url{https://doi.org} (Accessed 22 April 2026)}
}

@article{riehle2017optical,
    title = {Optical clock networks},
    author = {Riehle, Fritz and others},
    journal = {Nature Photonics},
    volume = {11},
    number = {1},
    pages = {25--31},
    year = {2017},
    doi = {10.1038/nphoton.2016.235},
    url = {https://doi.org/10.1038/nphoton.2016.235}
}

@article{Barontini2025DetectingMilliHzGWs,
  doi = {10.1088/1361-6382/ae09ec},
  url = {https://doi.org/10.1088/1361-6382/ae09ec},
  year = {2025},
  month = {oct},
  publisher = {IOP Publishing},
  volume = {42},
  number = {20},
  pages = {20LT01},
  author = {Barontini, Giovanni and others},
  title = {Detecting {milli-Hz} gravitational waves with optical resonators},
  journal = {Classical and Quantum Gravity}
  }

@article{Zhu2025TransportableApplications,
  author       = {Xian-Qing Zhu and others},
  title        = {Transportable single-crystal silicon ultra-stable cavity toward space applications},
  journal      = {Optica},
  volume       = {12},
  number       = {9},
  pages        = {1342--1349},
  year         = {2025},
  doi          = {10.1364/OPTICA.568436},
}

@article{Braxmaier2002CryogenicOpticalResonator,
  author       = {Claus Braxmaier and others},
  title        = {Tests of relativity using a cryogenic optical resonator},
  journal      = {Phys. Rev. Lett.},
  volume       = {88},
  number       = {1},
  pages        = {010401},
  year         = {2002},
  doi          = {10.1103/PhysRevLett.88.010401},
  url          = {https://doi.org/10.1103/PhysRevLett.88.010401},
}

@article{Toksoz1974,
  author  = {Toks{\"o}z, M. Nafi and others},
  title   = {Structure of the Moon},
  journal = {Reviews of Geophysics and Space Physics},
  volume  = {12},
  number  = {4},
  pages   = {539--567},
  year    = {1974},
  doi     = {10.1029/RG012i004p00539}
}

@article{Sanjuan2019LongTermStableOpticalCavity,
  author       = {Josep Sanjuan and others},
  title        = {Long-term stable optical cavity for special relativity tests in space},
  journal      = {Opt. Express},
  volume       = {27},
  number       = {25},
  pages        = {36206--36220},
  year         = {2019},
  doi          = {10.1364/OE.27.036206},
  url          = {https://doi.org/10.1364/OE.27.036206}
}

@unpublished{Lewis2026,
  author = {Ben Lewis and others},
  title  = {Towards a $1\times10^{-17}$ optical cavity},
  year = {2026},
  note   = {Proceedings of the International Frequency Control Symposium, in press}
}

@misc{NASA_visualization,
  author       = {{NASA Scientific Visualization Studio} and others},
  title        = {Lunar south pole illumination with {E}arth and {S}un},
  year  = {2024},
  howpublished = {\url{https://svs.gsfc.nasa.gov/5228/}},
  note         = {Visualizer: Ernie Wright},
  organization = {NASA Goddard Space Flight Center}
}

@article{Edlen1966RefractiveIndexAir,
  author       = {Bengt Edlén and others},
  title        = {The refractive index of air},
  journal      = {Metrologia},
  volume       = {2},
  number       = {2},
  pages        = {71--80},
  year         = {1966},
  doi          = {10.1088/0026-1394/2/2/002}
}

@article{nunn2020lunar2,
  title = {Lunar seismology: a data and instrumentation review},
  author = {Nunn, C. and others},
  journal = {Space Sci. Rev.},
  doi = {10.1007/s11214-020-00709-3},
  volume = {216},
  pages = {89},
  year = {2020}
}

@article{nunn2020lunar,
  title = {Lunar seismology: a data and instrumentation review},
  author = {Nunn, C. and others},
  journal = {Space Sci. Rev.},
  doi = {10.1007/s11214-020-00709-3},
  volume = {216},
  pages = {89},
  year = {2020}
}

@article{Ashby2024AMoon,
    title = {{A relativistic framework to estimate clock rates on the Moon}},
    year = {2024},
    journal = {Astronom. J.},
    author = {Ashby, Neil and others},
    number = {3},
    month = {9},
    pages = {112},
    volume = {168},
    publisher = {American Astronomical Society},
    url = {https://iopscience.iop.org/article/10.3847/1538-3881/ad643a},
    doi = {10.3847/1538-3881/ad643a},
    issn = {0004-6256}
}

@inproceedings{Latham1970MoonquakesMeteoroidsLunarInterior,
  author    = {G. Latham and others},
  title     = {Moonquakes, meteoroids, and the state of the lunar interior},
  booktitle = {Lunar and Planetary Science Conference Proceedings},
  year      = {1973}
}

@article{Watkins1972ApolloExperiment,
  author  = {Watkins, J. S. and Kovach, R. L.},
  title   = {Apollo 14 active seismic experiment},
  journal = {Science},
  volume  = {175},
  pages   = {1244--1245},
  year    = {1972}
}

@article{Burt2021DemonstrationSpace,
    title = {{Demonstration of a trapped-ion atomic clock in space}},
    year = {2021},
    journal = {Nature},
    author = {Burt, E. A. and others},
    number = {7865},
    month = {6},
    pages = {43--47},
    volume = {595},
    publisher = {Nature Publishing Group},
    url = {https://www.nature.com/articles/s41586-021-03571-7},
    doi = {10.1038/s41586-021-03571-7},
    issn = {1476-4687},
    pmid = {34194022}
}

@article{Paige2010DivinerRegion,
    title = {{Diviner lunar radiometer observations of cold traps in the moon's south polar region}},
    year = {2010},
    journal = {Science},
    author = {Paige, David A. and others},
    number = {6003},
    month = {10},
    pages = {479--482},
    volume = {330},
    doi = {10.1126/science.1187726},
    issn = {00368075},
    pmid = {20966246}
}

@article{Lorenz2018EmpiricalSurfaces,
    title = {{Empirical recurrence rates for ground motion signals on planetary surfaces}},
    year = {2018},
    journal = {Icarus},
    author = {Lorenz, Ralph D. and others},
    month = {3},
    pages = {273--279},
    volume = {303},
    url = {https://linkinghub.elsevier.com/retrieve/pii/S0019103517302178},
    doi = {10.1016/j.icarus.2017.10.008},
    issn = {00191035}
}

@article{Siegler2015EvolutionStability,
    title = {{Evolution of lunar polar ice stability}},
    year = {2015},
    journal = {Icarus},
    author = {Siegler, Matt and others},
    month = {7},
    pages = {78--87},
    volume = {255},
    publisher = {Academic Press Inc.},
    doi = {10.1016/j.icarus.2014.09.037},
    issn = {10902643}
}

@article{Bussey1999IlluminationPole,
    title = {{Illumination conditions at the lunar south pole}},
    year = {1999},
    journal = {Geophys. Res. Lett.},
    author = {Bussey, D. Ben J. and others},
    number = {9},
    month = {5},
    pages = {1187--1190},
    volume = {26},
    publisher = {American Geophysical Union},
    doi = {10.1029/1999GL900213},
    issn = {00948276}
}

@ARTICLE{Bourgoin2025LunarTimescale,
  author = {Bourgoin, A and others},
  title = {Lunar reference timescale},
  journal = {Metrologia},
  year = {2026},
  volume = {63},
  pages = {015003},
  number = {1},
  month = {jan},
  doi = {10.1088/1681-7575/ae2c03},
  publisher = {IOP Publishing}
}

@article{Kopeikin2024LunarRelativity,
    title = {{Lunar time in general relativity}},
    year = {2024},
    journal = {Phys. Rev. D},
    author = {Kopeikin, Sergei M. and others},
    number = {8},
    month = {10},
    pages = {084047},
    volume = {110},
    publisher = {American Physical Society},
    url = {https://journals.aps.org/prd/abstract/10.1103/PhysRevD.110.084047},
    doi = {10.1103/PhysRevD.110.084047},
    issn = {24700029},
    arxivId = {2407.04862}
}

@article{Majstorovic2025ModelingForcing,
    title = {{Modeling lunar response to gravitational waves using normal-mode approach and tidal forcing}},
    year = {2025},
    journal = {Phys. Rev. D},
    author = {Majstorovi{\'{c}}, Josipa and others},
    number = {4},
    month = {2},
    pages = {044061},
    volume = {111},
    publisher = {American Physical Society},
    url = {https://journals.aps.org/prd/abstract/10.1103/PhysRevD.111.044061},
    doi = {10.1103/PhysRevD.111.044061},
    issn = {24700029},
    arxivId = {2411.09559}
}

@article{Bickel2021PeeringLearning,
    title = {{Peering into lunar permanently shadowed regions with deep learning}},
    year = {2021},
    journal = {Nature Commun.},
    author = {Bickel, V. T. and others},
    number = {1},
    month = {12},
    volume = {12},
    pages = {5607},
    publisher = {Nature Communs.},
    doi = {10.1038/s41467-021-25882-z},
    issn = {20411723},
    pmid = {34556656}
}

@article{BenBussey2003PermanentPoles,
    title = {{Permanent shadow in simple craters near the lunar poles}},
    year = {2003},
    journal = {Geophys. Res. Lett.},
    author = {Ben Bussey, D J and others},
    number = {6},
    month = {3},
    pages = {1278},
    volume = {30},
    publisher = {John Wiley {\&} Sons, Ltd},
    url = {https://onlinelibrary.wiley.com/doi/full/10.1029/2002GL016180},
    doi = {10.1029/2002GL016180},
    issn = {1944-8007}
}

@article{Williams2019SeasonalMoon,
  author  = {Williams, J.-P. and others},
  title   = {Seasonal polar temperatures on the moon},
  journal = {J. Geophys. Res. Planets},
  volume  = {124},
  pages   = {2505--2521},
  year    = {2019}
}

@article{Williams2019SeasonalMoon2,
  author  = {Williams, J.-P. and others},
  title   = {Seasonal polar temperatures on the moon},
  journal = {J. Geophys. Res. Planets},
  volume  = {124},
  pages   = {2505--2521},
  year    = {2019}
}

@article{Lezius2016Space-borneMetrology,
    title = {{Space-borne frequency comb metrology}},
    year = {2016},
    journal = {Optica},
    author = {Lezius, Matthias and others},
    number = {12},
    month = {12},
    pages = {1381-1387},
    volume = {3},
    publisher = {Optica Publishing Group},
    doi = {10.1364/optica.3.001381},
    issn = {23342536}
}

@article{Duennebier1974ThermalMoonquakes,
    title = {{Thermal moonquakes}},
    year = {1974},
    journal = {J. Geophys. Res.},
    author = {Duennebier, Frederick and others},
    number = {29},
    month = {10},
    pages = {4351--4363},
    volume = {79},
    publisher = {John Wiley {\&} Sons, Ltd},
    url = {https://onlinelibrary.wiley.com/doi/full/10.1029/JB079i029p04351 https://onlinelibrary.wiley.com/doi/abs/10.1029/JB079i029p04351 https://agupubs.onlinelibrary.wiley.com/doi/10.1029/JB079i029p04351},
    doi = {10.1029/JB079I029P04351},
    issn = {2156-2202}
}

@article{Johnson1972VacuumSurface,
    title = {{Vacuum measurements on the lunar surface}},
    year = {1972},
    journal = {J. Vac. Sci. Technol.},
    author = {Johnson, Francis S. and others},
    number = {1},
    month = {1},
    pages = {450--456},
    volume = {9},
    publisher = {American Vacuum Society},
    doi = {10.1116/1.1316652},
    issn = {0022-5355}
}

@article{JWST2022,
    title = {Conceptual Design and Development History of the {MIRI} Cryocooler System on {JWST}},
    year = {2022},
    journal = {Cryocoolers},
    author = {Ross, R. G. and others},
    pages = {1--20},
    volume = {22}}

@article{Menzel_2023,
    doi = {10.1088/1538-3873/acbb9f},
    url = {https://doi.org/10.1088/1538-3873/acbb9f},
    year = {2023},
    month = {jun},
    publisher = {The Astronomical Society of the Pacific},
    volume = {135},
    number = {1047},
    pages = {058002},
    author = {Menzel, M. and others},
    title = {The Design, Verification, and Performance of the {J}ames {W}ebb Space Telescope},
    _journal = {Publications of the Astronomical Society of the Pacific},
    journal = {Publ. Astron. Soc. Pac.},
    abstract = {The James Webb Space Telescope (JWST) is NASA’s flagship mission successor to the highly successful Hubble Space Telescope. It is an infrared observatory featuring a cryogenic 6.6 m aperture, deployable Optical Telescope Element (OTE) with a payload of four science instruments (SIs) assembled into an Integrated Science Instrument Module (ISIM) that provide imagery and spectroscopy in the near-infrared band between 0.6 and 5 μm and in the mid-infrared band between 5 and 28.1 μm. JWST was successfully launched on 2021 December 25 aboard an Ariane 5 launch vehicle. All 50 major deployments were successfully completed on 2022 January 8. The observatory performed all midcourse correction maneuvers and achieved its operational mission orbit around the Sun–Earth second Lagrange point (L2). All commissioning and calibration activities have been completed, and JWST has begun its science mission. This paper will provide a description of the driving requirements and their technical challenges, the engineering processes involved in the design formulation, the resulting observatory design, the verification programs that proved it to be flightworthy, and the measured on-orbit performance of the observatory. Since companion papers will describe the details of the OTE and SIs, this paper will concentrate on describing the key features of the observatory architecture that accommodates these elements, particularly those features and capabilities associated with accommodating the radiometric and image-quality performance.}
}

@article{Bugby2003,
  author  = {Bugby, D. and Marland, B. and Stouffer, C. and Kroliczek, E.},
  title   = {Advanced components and techniques for cryogenic integration},
  journal = {Cryocoolers},
  volume  = {12},
  pages   = {693--708},
  year    = {2003}
}

@article{Nunn2021,
    title = {Standing on {A}pollo’s Shoulders: A Microseismometer for the Moon.},
    year = {2021},
    journal = {Planet. Sci. J.},
    author = {Nunn, C and others},
    pages = {36},
    volume = {2},
    url = {https://doi.org/10.3847/PSJ/abd63b}
}

@article{Nakamura1982,
  author       = {Yosio Nakamura and others},
  title        = {Apollo lunar seismic experiment—final summary},
  journal      = {J. Geophys. Res.},
  volume       = {87},
  number       = {A1},
  pages        = {A117--A123},
  year         = {1982},
  doi          = {10.1029/JB087iS01p0A117}
}

@article{Latham1970,
  author       = {G. V. Latham and others},
  title        = {Passive seismic experiment},
  journal      = {Science},
  volume       = {167},
  number       = {3918},
  pages        = {455--457},
  year         = {1970},
  doi          = {10.1126/science.167.3918.455}
}

@misc{NASA_Apollo11Seismic,
  author       = {{NASA} and others},
  title        = {Apollo 11 Seismic Experiment},
  howpublished = {\url{https://science.nasa.gov/resource/apollo-11-seismic-experiment/}},
  year = {2017}
}

@article{Bothwell2022,
  title        = {Resolving the gravitational redshift across a millimetre-scale atomic sample},
  author       = {Bothwell, Tobias and others},
  journal      = {Nature},
  volume       = {602},
  pages        = {420--424},
  year         = {2022},
  doi          = {10.1038/s41586-021-04349-7}
}

@article{Lee2026,
  title        = {Frequency Stability of $2.5\times10^{-17}$ from a {Si} Cavity with {AlGaAs} Crystalline Mirrors},
  author       = {Lee, Dahyeon and others},
  journal      = {Phys. Rev. Lett.},
  volume       = {136},
  number       = {3},
  pages        = {033801},
  year         = {2026},
  doi          = {10.1103/PhysRevLett.136.033801},
}

@article{Oelker2019,
  title        = {Demonstration of $4.8\times10^{-17}$ stability at 1 s for two independent optical clocks},
  author       = {Oelker, Eric and others},
  journal      = {Nat. Photonics},
  volume       = {13},
  number       = {10},
  pages        = {714--719},
  year         = {2019},
  doi          = {10.1038/s41566-019-0493-4}
}

@article{Kolkowitz2016,
  title = {Gravitational wave detection with optical lattice atomic clocks},
  author = {Kolkowitz, S. and others},
  journal = {Phys. Rev. D},
  volume = {94},
  issue = {12},
  pages = {124043},
  numpages = {15},
  year = {2016},
  month = {Dec},
  publisher = {American Physical Society},
  doi = {10.1103/PhysRevD.94.124043},
  url = {https://link.aps.org/doi/10.1103/PhysRevD.94.124043}
}

@article{Mehlstaubler2018,
  title        = {Atomic clocks for geodesy},
  author       = {Mehlst{\"a}ubler, T. E. and others},
  journal      = {Rep. Prog. Phys.},
  volume       = {81},
  number       = {6},
  pages        = {064401},
  year         = {2018},
  doi          = {10.1088/1361-6633/aab409},
}

@Article{Pikovski2012,
author={Pikovski, Igor
and Vanner, Michael R.
and Aspelmeyer, Markus
and Kim, M. S.
and Brukner, {\v{C}}aslav and others},
title={Probing {P}lanck-scale physics with quantum optics},
journal={Nature Physics},
year={2012},
month={May},
day={01},
volume={8},
number={5},
pages={393-397},
issn={1745-2481},
doi={10.1038/nphys2262},
url={https://doi.org/10.1038/nphys2262}
}

@article{Turyshev2025,
  title        = {Lunar laser ranging with high-power continuous-wave lasers},
  author       = {Turyshev, Slava G. and others},
  journal      = {Phys. Rev. Appl.},
  volume       = {23},
  number       = {6},
  pages        = {064066},
  year         = {2025},
  doi          = {10.1103/PhysRevApplied.23.064066},
}

@article{Berceau2016,
  title        = {Space-time reference with an optical Llink},
  author       = {Berceau, Paul and others},
  journal      = {Class. Quantum Grav.},
  volume       = {33},
  number       = {13},
  pages        = {135007},
  year         = {2016},
  doi          = {10.1088/0264-9381/33/13/135007},
}

@article{Derevianko2022,
  title        = {Fundamental Physics with a State-of-the-Art Optical Clock in Space},
  author       = {Derevianko, Andrei and others},
  journal      = {Quantum Sci. Technol.},
  volume       = {7},
  number       = {4},
  pages        = {044002},
  year         = {2022},
  doi          = {10.1088/2058-9565/ac7df9},
}

@article{Milner2019OpticalTimescale,
  author       = {William R. Milner and others},
  title        = {Demonstration of a Time Scale Based on a Stable Optical Carrier},
  journal      = {Phys. Rev. Lett.},
  volume       = {123},
  number       = {17},
  pages        = {173201},
  year         = {2019},
  doi          = {10.1103/PhysRevLett.123.173201},
  url          = {https://doi.org/10.1103/PhysRevLett.123.173201}
}

@article{Yao2019OpticalClockTimeScale,
  author       = {Yao, Jian and others},
  title        = {Optical-Clock-Based Time Scale},
  journal      = {Phys. Rev. Appl.},
  volume       = {12},
  number       = {4},
  pages        = {044069},
  year         = {2019},
  doi          = {10.1103/PhysRevApplied.12.044069},
  url          = {https://doi.org/10.1103/PhysRevApplied.12.044069}
}

@article{Hachisu2018MonthsLongTimeScale,
  author       = {Hachisu, Hiroyuki and others},
  title        = {Months-long real-time generation of a time scale based on an optical clock},
  journal      = {Sci. Rep.},
  volume       = {8},
  number       = {1},
  pages        = {4243},
  year         = {2018},
  doi          = {10.1038/s41598-018-22423-5},
  url          = {https://doi.org/10.1038/s41598-018-22423-5}
}

@article{Formichella2024YearLongOpticalTimeScale,
  author       = {Formichella, V. and others},
  title        = {Year-long optical time scale with sub-nanosecond capabilities},
  journal      = {Optica},
  volume       = {11},
  number       = {4},
  pages        = {523--530},
  year         = {2024},
  doi          = {10.1364/OPTICA.497101},
  url          = {https://doi.org/10.1364/OPTICA.497101}
}

@article{Zhang2023SrOpticalTimeScaleNTSC,
  author  = {Zhang, Q. and others},
  title   = {Demonstration of a time scale with the $^{87}${Sr} optical lattice clock at {NTSC}},
  journal = {AIP Adv.},
  volume  = {13},
  pages   = {115316},
  year    = {2023}
}

@article{Ely2025BenefitSpaceClocksDSN,
  author       = {Ely, T. and others},
  title        = {The Benefit of Space Clocks for the Deep Space Network},
  journal      = {Radio Sci.},
  volume       = {60},
  number       = {8},
  year         = {2025},
  pages        = {e2025RS008244},
  doi          = {10.1029/2025RS008244},
  url          = {https://doi.org/10.1029/2025RS008244}
}

@inproceedings{FFS_2024,
  author = {Aboobaker, A. and others},
  title = {The Farside Seismic Suite: A Novel Approach for Long-term Lunar Seismology},
  booktitle = {2024 IEEE Aerospace Conference},
  year = {2024},
  doi = {10.1109/AERO58975.2024.10521223},
  pages = {1--8},
  address = {Big Sky, USA},
}

@inproceedings{lunanet2020,
  author    = {Israel, D. J. and others},
  title     = {{LunaNet}: A flexible and extensible lunar exploration communications and navigation infrastructure},
  booktitle = {2020 IEEE Aerospace Conference},
  pages     = {1--14},
  year      = {2020}
}

@ARTICLE{Wcislo_Dark_Matter_2016,
  author = {P. Wcis{\l}o and others},
  title = {Experimental constraint on dark matter detection with optical atomic
	clocks},
  journal = {Nat. Astron.},
  year = {2016},
  volume = {1},
  pages = {0009},
  doi = {10.1038/s41550-016-0009},
}

@article{Battat_2009,
    doi = {10.1086/596748},
    url = {https://doi.org/10.1086/596748},
    year = {2009},
    month = {jan},
    publisher = {University of Chicago Press},
    volume = {121},
    number = {875},
    pages = {29},
    author = {Battat, J. B. R. and others},
    title = {The {A}pache {P}oint Observatory Lunar Laser-ranging Operation ({APOLLO}): Two Years of Millimeter-Precision Measurements of the Earth-Moon Range},
    _journal = {Publications of the Astronomical Society of the Pacific},
    journal = {Publ. Astron. Soc. Pac.},
    abstract = {In 2006 April, the Apache Point Observatory Lunar Laser-ranging Operation (APOLLO) began its science campaign to measure the Earth-Moon separation to millimeter precision. Since that time more than 280 “normal-point” measurements have been made of the distance between the Apache Point Observatory (APO) 3.5-m telescope in New Mexico and retro-reflector arrays on the surface of the Moon. If only statistical errors are considered, then the median nightly range measurement uncertainty for all of our data is 1.8 mm of one-way path, and is 1.1 mm for data after 2007 September. We present an analysis of the APOLLO system performance, highlighting the record-breaking photon return rates and the ability to perform high-cadence observations of multiple lunar retro-reflector targets in a short (30–60 minute) time span. We also show that there is no evidence to suggest that the APOLLO apparatus introduces drifts in the lunar-range measurement over timescales of minutes to an hour.}
}

@article{Shen_OpticalTimeTransfer,
author = {Qi Shen and others},
journal = {Optica},
number = {4},
pages = {471--476},
publisher = {Optica Publishing Group},
title = {Experimental simulation of time and frequency transfer via an optical satellite--ground link at 10-18 instability},
volume = {8},
month = {Apr},
year = {2021},
url = {https://opg.optica.org/optica/abstract.cfm?URI=optica-8-4-471},
doi = {10.1364/OPTICA.413114}
}

@article{Caldwell_OpticalTimeTransfer,
author = {Emily D. Caldwell and others},
journal = {Adv. Opt. Photon.},
number = {2},
pages = {375--440},
publisher = {Optica Publishing Group},
title = {High-precision optical time and frequency transfer},
volume = {17},
month = {Jun},
year = {2025},
url = {https://opg.optica.org/aop/abstract.cfm?URI=aop-17-2-375},
doi = {10.1364/AOP.545290}
}
\end{document}